\documentclass[twocolumn,superscriptaddress,aps,preprintnumbers,amsmath,amssymb,prl,nofootinbib]{revtex4-2}

\usepackage{dcolumn}% Align table columns on the decimal point
\usepackage[dvipdfmx]{graphicx}
\usepackage{epstopdf}
\usepackage{bm}% bold math
\usepackage{color}
\usepackage{mathrsfs}
\usepackage{physics}
\usepackage{comment}
\usepackage{here}

\newcommand{\be}{\begin{equation}}  
\newcommand{\ee}{\end{equation}}
\newcommand{\ba}{\begin{eqnarray}}
\newcommand{\ea}{\end{eqnarray}}

\newcommand{\rd}{{\rm d}}

\def\Mpl{M_{\rm 
Pl}}

\def\gc{g_{a \gamma \gamma}}

\definecolor{Tomato}{cmyk}{0.37, 1.00, 1.00, 0.03}

\begin{document}

\preprint{YITP-24-19, WUCG-24-01}

\title{
Probing the signature of axions through the quasinormal modes of black holes
}

\author{Antonio De Felice}
\email{antonio.defelice@yukawa.kyoto-u.ac.jp}
\affiliation{Center for Gravitational Physics and Quantum Information, 
Yukawa Institute for Theoretical Physics, Kyoto University, 
606-8502, Kyoto, Japan}

\author{Shinji Tsujikawa}
\email{tsujikawa@waseda.jp}
\affiliation{Department of Physics, Waseda University, 
3-4-1 Okubo, Shinjuku, Tokyo 169-8555, Japan}

\begin{abstract}
The axion-photon coupling allows the existence of a magnetically and electrically charged black hole (BH) solution endowed with a pseudo-scalar hair. 
For the Reissner-Nordstr\" om BH with a given total charge and mass, it is known that the quasinormal modes (QNMs) are independent of the mixture between the magnetic and electric charges due to the presence of electric-magnetic duality.
We show that the BH with an axion hair breaks this degeneracy by realizing nontrivial QNMs that depend on the ratio between the magnetic and total charges.
Thus, the upcoming observations of BH QNMs through gravitational waves offer an exciting possibility for probing the existence of both magnetic monopoles and the axion coupled to photons.
\end{abstract}

\maketitle
%%%%%%%%%%%%%%%%%%%%%%%%%%%%%%%%%%%%%%%%%%%%%%%%%%%%%%%%%%%%%%%

%=============================================%
\section{I. Introduction}
%=============================================%

The advent of gravitational-wave astronomy opened up a new window for probing the physics in strong-gravity regimes \cite{LIGOScientific:2016aoc}. 
From the merger events of compact binaries, one can constrain not only the masses and charges of black holes (BHs) but also quasinormal modes (QNMs) of damped oscillations.
QNMs of the Schwarzschild BH can be modified by the presence of extra degrees of freedom \cite{Kokkotas:1999bd, Nollert:1999ji, Berti:2009kk, Konoplya:2011qq, Pani:2013pma, Guo:2022rms}, e.g., vector and scalar fields. 
A simple example is the Reissner-Nordstr\" om (RN) BH 
with an electric charge \cite{Gunter:1980,Kokkotas:1988fm,Leaver:1990zz,Berti:2003zu},
which arises from the presence of 
a vector field $A_{\mu}$ in Einstein-Maxwell theory.

Recently, there has been growing interest in 
understanding properties of the magnetically 
charged BHs \cite{Maldacena:2020skw}. 
Such BHs may have primordial origins as a 
result of the absorption of magnetic monopoles
in the early Universe \cite{Stojkovic:2004hz, Kobayashi:2021des, Das:2021wei, Estes:2022buj, Zhang:2023tfv}. 
Since the magnetic BH is not neutralized with 
ordinary matter in conductive media, it can be 
a more stable configuration relative to 
the purely electric BHs \cite{Maldacena:2020skw, Bai:2020ezy}. 
Then, it is worth studying observational signatures of the magnetic monopole carried by BHs. 
With a given total BH charge and mass, however, it was recently shown that the QNM of the RN BH is the same independent of the mixture between the magnetic and electric charges \cite{DeFelice:2023rra} (see also Refs.~\cite{Nomura:2020tpc, Nomura:2021efi, Pereniguez:2023wxf}).
Hence we cannot distinguish between the magnetic and electric RN BHs from the observations of QNMs. 

In the presence of an additional scalar field, it is possible to realize nontrivial BH solutions endowed with scalar hairs.
The pseudo-scalar axion field $\phi$, which was originally 
introduced to address the strong CP problem 
in QCD \cite{Peccei:1977hh}, can be coupled to an 
electromagnetic field strength tensor $F_{\mu \nu}$ 
in the form $-(1/4)\gc \phi F_{\mu \nu}\tilde{F}^{\mu \nu}$, 
where $\gc$ is a coupling constant and $\tilde{F}^{\mu \nu}$ 
is a dual of $F^{\mu \nu}$. 
In string theory, there are also axion-like light 
particles with a vast range of masses \cite{Svrcek:2006yi}.
It is known that there are BHs endowed with the axion hair 
as well as with the magnetic and 
electric charges \cite{Lee:1991jw, Filippini:2019cqk, Fernandes:2019kmh}.
An important question is whether or not such hairy BHs can be observationally distinguished from the RN BH.

In this letter, we compute the QNMs of hairy BHs in Einstein-Maxwell-axion (EMA) theory in the presence of the axion-photon coupling.
We show that, with a given total BH charge and mass, the QNMs are different depending on the ratio between the magnetic and electric charges. This property is in stark contrast with that of the RN BH. 
Thus, the precise observations of QNMs can allow us to probe the existence of both the magnetic monopole and the axion.

%=============================================%
\section{II. Hairy BHs in EMA theory}
%=============================================%

The EMA theory is given by the action 
\ba   
{\cal S}&=&\int {\rm d}^{4}x \sqrt{-g} 
\left[ \frac{\Mpl^2}{2} R-\frac{1}{4}F_{\mu \nu}F^{\mu \nu}
-\frac{1}{2} g^{\mu\nu} \nabla_{\mu}\phi 
\nabla_{\nu}\phi\right.\nonumber\\
& & \left.-\frac{1}{2}m_{\phi}^2 \phi^2-\frac{1}{4}
\gc \phi F_{\mu \nu}\tilde{F}^{\mu \nu}
\right]\,,
\label{action}
\ea
where $g$ is the determinant of metric tensor $g_{\mu\nu}$, $\Mpl$ is the reduced Planck mass, $R$ is the Ricci scalar, and $m_{\phi}$ is the axion mass. 
The field strength tensor $F_{\mu \nu}$ is related to the vector field 
$A_{\mu}$, as $F_{\mu \nu}=\nabla_{\mu} A_{\nu}-\nabla_{\nu} A_{\mu}$, and $\tilde{F}^{\mu \nu}=\epsilon^{\mu \nu \rho \sigma} F_{\rho \sigma}/(2\sqrt{-g})$ with $\epsilon^{0123}=+1$. 
The action (\ref{action}) respects $U(1)$ gauge invariance under 
the shift $A_{\mu} \to A_{\mu}+\nabla_{\mu} \chi$.

We consider a static and spherically symmetric line element given by 
\be
{\rm d}s^{2} =-f(r) {\rm d}t^{2} +h^{-1}(r) {\rm d}r^{2} + 
r^{2} \left( {\rm d}\theta^{2}
+\sin^{2}\theta\, {\rm d} \varphi^{2} \right)\,,
\label{spmetric}
\ee
where $f$ and $h$ are functions of the radial coordinate $r$. The axion and vector-field configurations compatible with this background are $\phi=\phi(r)$ and $A_{\mu}=[ A_0(r), 0, 0, -q_M \cos \theta]$, where $q_M$ is a constant corresponding to the magnetic charge.
The axion and the temporal vector component obey the following differential equations 
\ba
\hspace{-0.4cm}
& &
\phi''+\left( \frac{2}{r}+\frac{f'}{2f} +\frac{h'}{2h} \right)\! \phi'-\frac{m_{\phi}^2}{h}\phi
-\frac{\gc q_M A_0'}{r^2 \sqrt{fh}}=0\,,
\label{eq:eqphi}\\
\hspace{-0.4cm}
& &
A_0'=\frac{\sqrt{f} [q_E+q_M\gc \phi]}{r^2 \sqrt{h}}\,,
\label{A0r}
\label{eq:A0}
\ea
respectively, where a prime represents the derivative with respect to $r$. The integration constant $q_E$ in $A_0'$ corresponds to the electric charge. 
For $q_M \neq 0$, the BH can have a nontrivial axion profile through the coupling with $A_0'$.
The gravitational equations of motion are given by 
\ba 
\hspace{-0.8cm}
& &
\frac{r h'+h-1}{\Mpl^{-2} h}
+\frac{r^2 \phi'^2}{2}+\frac{r^2 m_{\phi}^2\phi^2}{2h}
+\frac{q_{M}^{2}}{2 h \,r^{2}}
+\frac{r^{2}A_{0}'^{2}}{2f}=0\,,
\label{eq:h}\\
\hspace{-0.8cm}
& & \Delta \equiv r_h \left( \frac{f'}{f}-\frac{h'}{h}
\right) = \frac{r_h r \phi'^2}{\Mpl^2}\,,
\label{dfh}
\ea
where $r_h$ is the outer horizon radius. 
Around $r=r_h$, we expand the metrics and scalar field, as 
\ba
& &
f=\sum_{i=1} f_i (r-r_h)^i\,,\qquad 
h=\sum_{i=1} h_i (r-r_h)^i\,, \nonumber \\
& &
\phi=\phi_0+\sum_{i=1} \phi_i (r-r_h)^i\,,
\label{fhexpan}
\ea
where $f_i$, $h_i$, $\phi_0$ and $\phi_i$ are constants. 
For consistency with the background equations, we require that 
\ba
h_1 &=& [2 \Mpl^2 r_h^2-q_E^2-q_M^2
-\gc q_M \phi_0 (\gc q_M \phi_0+2q_E) \nonumber \\
& &
-m_{\phi}^2 r_h^4 \phi_0^2]/(2 \Mpl^2 r_h^3)\,,
\label{h1}\\
\phi_1 &=& [(m_{\phi}^{2} r_{h}^{4}+\gc^{2} q_{M}^{2}) 
\phi_{0}+\gc  q_{M} q_{E}]/(h_{1} r_{h}^{4})\,.
\label{phi1}
\ea
We are interested in hairy BH solutions where $|\phi|$ is a decreasing function of $r$ from the horizon to spatial infinity. 
Furthermore, to ensure the property $h(r)>0$ for $r>r_h$, we require that $h_1>0$. 
Hence, around $r=r_h$, these two conditions lead to
\be
\phi_0 \phi_1 h_1 r_h^4=\left(m_{\phi}^{2} r_{h}^{4}
+\gc^{2} q_{M}^{2}\right) 
\phi_{0}^2+\gc \phi_0 q_{M} q_{E}<0\,.
\label{con1}
\ee
Then, it is at least necessary to satisfy the inequality  
\be
\gc \phi_0 q_{M} q_{E}<0\,.
\label{con2}
\ee
Since this condition is violated for $q_M=0$ or $q_E=0$, we need the existence of both magnetic and electric charges to realize 
a nontrivial axion hair. 
The inequality (\ref{con2}) does not hold for $\gc=0$ either, 
so we require the axion-photon coupling 
$-(1/4) \gc \phi F_{\mu \nu}\tilde{F}^{\mu \nu}$ to realize 
hairy BH solutions. In other words, the no-hair property of BHs 
for a canonical scalar field \cite{Hawking:1972qk,Bekenstein:1972ny} is broken by the appearance of 
a secondary axion hair through interaction with 
electromagnetic fields. 
 
Without loss of generality, we will consider the case $\phi_0>0$, $q_{M}>0$, $q_E>0$, and $\gc<0$.
Because of Eq.~\eqref{h1}, combining (\ref{con1}) with $h_1>0$ gives  
\be
\frac{-\gc q_M q_E-\sqrt{\cal A}}
{q_M^2 \gc^2+m_{\phi}^2 r_h^4} 
< \phi_0 <\frac{-\gc q_M q_E}
{q_M^2 \gc^2+m_{\phi}^2 r_h^4}\,,
\label{con3}
\ee
where 
\be
{\cal A} \equiv (2\Mpl^2 r_h^2-q_M^2) q_M^2 \gc^2+
\left( 2\Mpl^2 r_h^2-q_M^2-q_E^2 \right)
m_{\phi}^2r_h^4\,.
\ee
If $q_M^2 \geq 2\Mpl^2 r_h^2$, then ${\cal A}$ is negative. The magnetic charge should be at least in the range $q_M^2<2 \Mpl^2 r_h^2$ for the existence of hairy BHs with $\phi_0 \neq 0$. 
More strictly, so long as the condition 
\be
q_M^2+q_E^2<2\Mpl^2 r_h^2
\label{con4}
\ee
is satisfied, we always have ${\cal A}>0$ and hence there is the field value $\phi_0$ in the range (\ref{con3}). 

We search for the solutions respecting the asymptotic flatness, i.e., $f \to 1$, $h \to 1$, $f' \to 0$, and $h' \to 0$ as $r \to \infty$. 
We also impose the boundary condition $\phi(r \to \infty)=0$.
In this large-distance regime, Eq.~(\ref{eq:eqphi}) approximately reduces to $\phi''+2\phi'/r-m_{\phi}^2 \phi\simeq \gc q_M q_E/r^4$.
The solution to this equation, which respects the boundary condition $\phi(r \to \infty)=0$, can be expressed as
\be
\phi(r) \simeq q_s \frac{e^{-m_{\phi} r}}{r}
-\frac{\gc q_M q_E}{m_\phi^2 r^4}\,,
\label{philarge}
\ee
where $q_s$ is a constant.
The first term in Eq.~(\ref{philarge}) decreases exponentially for $r>m_{\phi}^{-1}$ and hence $\phi(r) \propto r^{-4}$ in this regime. 
For $m_{\phi}=0$, the large-distance solution is given by $\phi(r) \simeq q_s/r+\gc q_M q_E/(2r^2)$. 
In both cases, the metric approaches that of the RN BH as $r \to \infty$.

%%%%%%%%%%%%%%%%%%%%%%%%%%%%%%%%%%%%%%%%%%%%%%%%%%%
\begin{figure}[hb]
\centering
\includegraphics[height=2.5in,width=3.3in]{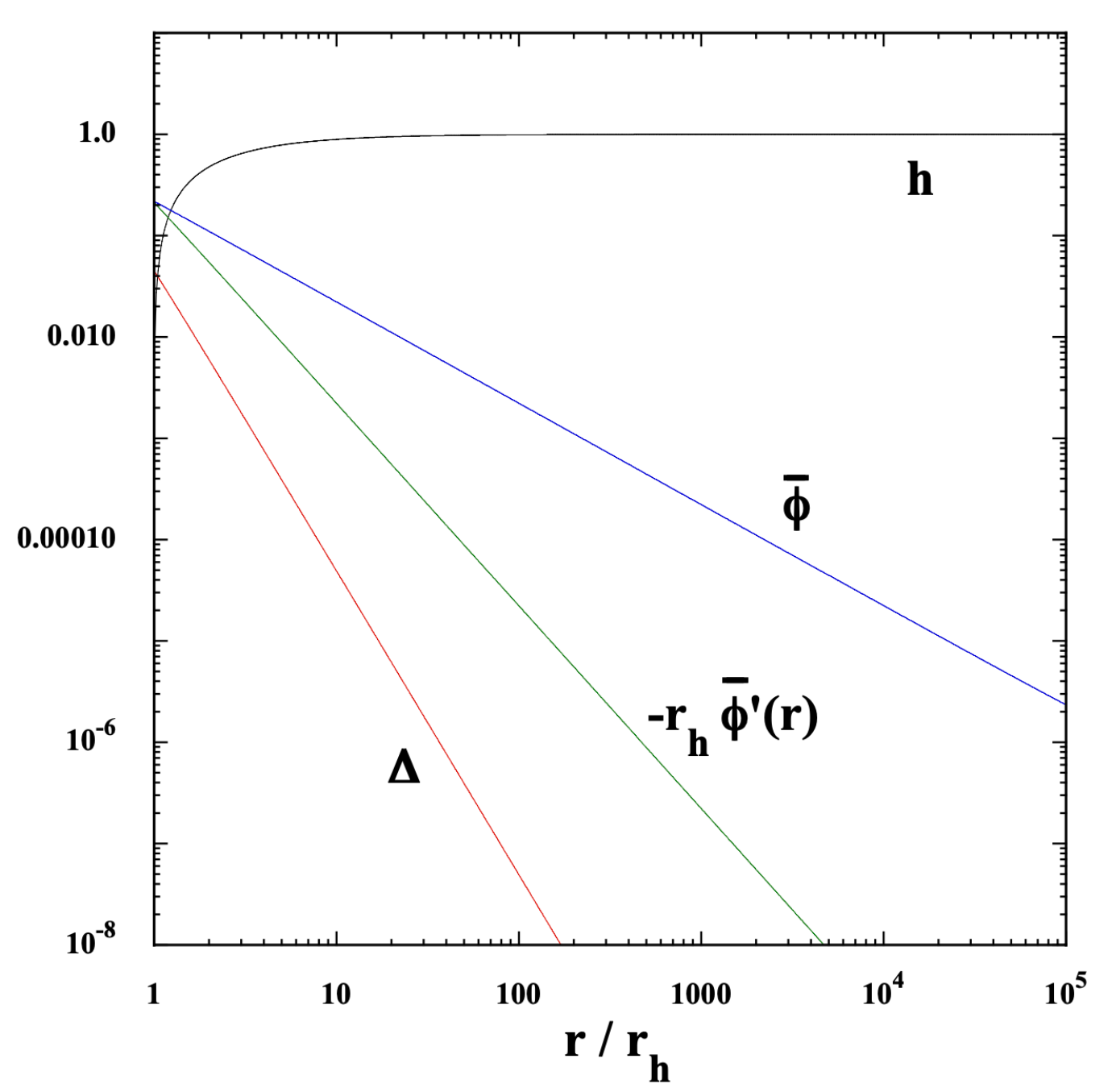}
\caption{We show $h$, $\Delta$, $\bar{\phi}$, 
and $-r_h \bar{\phi}'(r)$ versus $r/r_h$ for $m_{\phi}=0$, 
$\gc \Mpl=-10$, $q_M=0.05 \Mpl r_h$, 
and $q_E=0.5 \Mpl r_h$ with the field value 
$\phi_0=0.217899 \Mpl$ on the horizon.} 
\label{fig1}	
\end{figure}
%%%%%%%%%%%%%%%%%%%%%%%%%%%%%%%%%%%%%%%%%%%%%%%%%%%

To confirm the existence of hairy BH solutions, we numerically solve Eqs.~(\ref{eq:eqphi})-(\ref{dfh}) by imposing the aforementioned boundary conditions around $r=r_h$. In Fig.~\ref{fig1}, we plot $h$, $\Delta$, $\bar{\phi}=\phi/\Mpl$, and $-r_h \bar{\phi}'(r)$ versus $r/r_h$ for $m_{\phi}=0$, 
$\gc \Mpl=-10$, $q_M=0.05 \Mpl r_h$, and $q_E=0.5 \Mpl r_h$.
In this case, the two conditions (\ref{con2}) and (\ref{con4}) are satisfied, with $-1.827<\phi_0/\Mpl<1$ from Eq.~(\ref{con3}). 
The axion has a maximum amplitude $\phi_0 \simeq 0.217899 \Mpl$ on the horizon and then it decreases toward the asymptotic value $\phi(\infty)=0$ without changing the sign.
In Fig.~\ref{fig1}, we observe the field dependence $\phi'(r) =-q_s/r^2~(<0)$ in the regime $r \gg r_h$.
Substituting the large-distance solution $\phi(r)=q_s/r$ into Eqs.~(\ref{eq:h}) and (\ref{dfh}), we obtain 
$f=1-2M/r+(q_M^2+q_E^2)/(2\Mpl^2 r^2)+{\cal O}(r^{-3})$ and $h=f+q_s^2/(2\Mpl^2 r^2)+{\cal O}(r^{-3})$, with $\Delta \simeq r_h q_s^2/(\Mpl^2 r^3)$, where $M$ corresponds to the BH ADM mass.
As we see in Fig.~\ref{fig1}, the difference between $f'/f$ and $h'/h$ is most significant around $r=r_h$.

For $m_{\phi} \neq 0$ the axion has a growing-mode solution $e^{m_{\phi}r}/r$ manifesting at the distance $r \gtrsim 1/m_{\phi}$, but there should be appropriate boundary conditions respecting the regularities of both infinity and the horizon. 
We numerically confirm the existence of 
asymptotically-flat hairy BHs especially 
in the mass range $m_{\phi}r_h \lesssim 1$.
For a BH with $r_h \simeq 10^4$~m, the axion mass corresponding to $m_{\phi}r_h \lesssim 1$ is $m_{\phi} \lesssim 10^{-11}$~eV, which includes the case of fuzzy dark matter ($m_{\phi} \simeq 10^{-21}$~eV) \cite{Hu:2000ke}.
Taking the limit $m_{\phi} \to \infty$ in Eq.~(\ref{con3}), the allowed values of $\phi_0$ shrink to 0. Hence the hairy BH solution tends to disappear in this massive limit. 
For the axion mass $m_{\phi} \lesssim 1$~eV, the current limit on the axion-photon coupling is $|\gc| \lesssim 10^6 \Mpl^{-1}$ \cite{AxionLimits}.
The coupling $\gc$ used in Fig.~\ref{fig1} is well consistent with such a bound.

\section{III.~Quasinormal modes}

In this section, we compute the QNMs of hairy BHs in EMA theory by considering linear perturbations on the background (\ref{spmetric}). 
We choose the gauge in which the $\theta$ and $\varphi$ components 
of $h_{\mu \nu}$ vanish, i.e., 
\ba
\hspace{-0.3cm}
& &
h_{tt}=f(r) H_0 (t,r) Y_{l}(\theta), \quad 
h_{tr}=H_1 (t,r) Y_{l}(\theta),\nonumber \\
\hspace{-0.3cm}
& &
h_{t \varphi}=-Q(t,r) (\sin \theta) 
Y_{l, \theta} (\theta),
\;\; 
h_{rr}=h^{-1}(r) H_2(t,r) Y_{l}(\theta),\nonumber \\
\hspace{-0.3cm}
& &
h_{r \theta}=h_1 (t,r)Y_{l, \theta}(\theta),
\quad
h_{r \varphi}=-W(t,r) (\sin \theta) Y_{l,\theta} (\theta), 
\nonumber \\
\hspace{-0.3cm}
& &
h_{\theta \theta}=0,\quad 
h_{\varphi \varphi}=0,\quad
h_{\theta \varphi}=0,
\label{hcom}
\ea
where $Y_{l}(\theta)$'s are the $m=0$ components of spherical harmonics $Y_{lm}(\theta, \varphi)$, and 
$Y_{l, \theta} (\theta) \equiv \rd Y_{l}(\theta)/\rd \theta$. 
In Eq.~(\ref{hcom}), we omit the summation of $Y_{l} (\theta)$ concerning the multipoles $l$. 
Note that the above gauge choice is different from 
the Regge-Wheeler-Zerilli gauge \cite{Regge:1957td, Zerilli:1970se, Zerilli:1970wzz}, but the former can also fix the residual gauge 
degrees of freedom completely  \cite{Kobayashi:2012kh,Kobayashi:2014wsa,Kase:2021mix,Kase:2023kvq}. 
Since our theory has $U(1)$ gauge symmetry, we can choose a gauge 
in which the $\theta$ component of the vector-field perturbation 
$\delta A_{\mu}$ vanishes \cite{Kase:2023kvq}. 
Then, we consider the perturbed components 
of vector and axion fields, as
\ba
& &
\delta A_t=\delta A_0 (t,r) Y_{l}(\theta),\quad 
\delta A_r=\delta A_1 (t,r) Y_{l}(\theta),\quad \nonumber \\
& &
\delta A_\theta=0,\quad 
\delta A_{\varphi}=-\delta A(t,r) (\sin \theta) 
Y_{l,\theta}(\theta),
\nonumber \\
& &
\delta \phi=\delta \phi (t,r) Y_l (\theta),
\label{perma}
\ea
respectively. The coupling $-(\sqrt{-g}/4)\gc \phi F_{\mu \nu}\tilde{F}^{\mu \nu}$ in the action (\ref{action}) does not have contributions from the metric components and depends 
only linearly on each perturbed field. 
Moreover, none of the perturbations are coupled 
to the modes with different values of $l$ or $m$. 
The background spherical symmetry allows us to set $m=0$ 
without loss of generality. 
Indeed, for fixed $l$, the second-order perturbed action does not 
depend on the values of $m$.
Since the perturbations in the odd- and even-parity sectors are mixed for $q_M \neq 0$, we must deal with them all at once.

For $l \geq 2$, we expand Eq.~(\ref{action}) up to quadratic order in perturbations. Then, the resulting second-order action contains ten perturbed variables 
$H_0$, $H_1$, $H_2$, $h_1$, $Q$, $W$, 
$\delta A_0$, $\delta A_1$, $\delta A$, and $\delta \phi$. 
The explicit form of the total second-order action ${\cal S}^{(2)}$ is given in Appendix~A.  
Introducing the new fields $\chi_1$, $v_1$, $\chi_2$ 
defined in Eqs.~(\ref{chi1}), (\ref{v1}), and (\ref{chi2}), 
respectively, we can express the action in terms of the five dynamical perturbations 
$\chi_1$, $v_1$, $\chi_2$, $\delta A$, $\delta \phi$, 
and their $t, r$ derivatives. 
The process for deriving the reduced second-order action is explained in Appendix B.
Here, $v_1$ and $\chi_1$ are associated with the even- and odd-parity gravitational perturbations, respectively, while $\chi_2$ and $\delta A$ arise from the vector-field perturbations in even- and odd-parity sectors, respectively. We also have the dynamical axion perturbation $\delta \phi$.

For the computational simplicity, we make the following field redefinitions 
\ba
\hspace{-0.3cm}
& &
\psi_1=\Mpl rh e^{i\omega t}v_1,\quad 
\psi_2=\Mpl r e^{i\omega t}\chi_1,\nonumber \\
\hspace{-0.3cm}
& &
\psi_3=r^2 e^{i\omega t}\chi_2,\quad
\psi_4=e^{i\omega t} \delta A\,,\quad 
\psi_5=r e^{i\omega t} \delta \phi\,,
\ea
where $\omega$ is an angular frequency. 
Introducing the tortoise coordinate 
$r_{*}=\int^{r} {\rm d}\tilde{r}/\sqrt{f(\tilde{r})\,h(\tilde{r})}$, 
the perturbation equations of motion can be schematically written as
\be
\frac{{\rm d}^{2}\psi_{i}}{{\rm d}r_{*}^{2}}
+B_{ij}\frac{{\rm d}\psi_{j}}{{\rm d}r_{*}}
+C_{ij}\psi_{j}=0\,,\quad i,j\in\{1,\dots,5\},
\label{psieq}
\ee
where the matrices $B_{ij}$ and $C_{ij}$ are background-dependent
quantities, and $C_{ij}$ also contain $\omega$.

On the horizon ($r_* \to -\infty$) and at spatial infinity ($r_* \to +\infty$), Eq.~(\ref{psieq}) is approximately given by ${\rm d}^{2}\psi_{i}/{{\rm d}r_{*}^{2}} 
\simeq -\omega^{2}\psi_{i}$. 
The QNMs are characterized by purely ingoing waves on the horizon and purely outgoing at spatial infinity, and hence
\be
\psi_i (r_* \to -\infty)=A_i e^{-i \omega r_*}\,,\quad 
\psi_i (r_* \to \infty)=B_i e^{+i \omega r_*}\,,
\ee
where $A_i$ and $B_i$ are constants. 
For the calculation of QNMs, we will exploit a matrix-valued direct integration method \cite{Pani:2013pma} based on higher-order expansions of $\psi_i$ both around the horizon and infinity. 
Using the background solutions (\ref{fhexpan}) expanded in the vicinity of the horizon, we have $r_* \simeq (f_1 h_1)^{-1/2}\ln (r/r_h-1)$ and hence the leading-order solutions to $\psi_i$ are proportional to $(r-r_h)^{-i \omega (f_1 h_1)^{-1/2}}$. 
Then, around $r=r_h$, we choose the following ansatz
\be
\psi_{i}^{\rm H}=(r-r_{h})^{-i\omega (f_1 h_1)^{-1/2}} 
\sum_{n=0} (\psi_{i}^{\rm H})^{(n)}\,(r-r_{h})^{n}\,,
\label{psiH}
\ee
where $(\psi_{i}^{\rm H})^{(n)}$ is the $n$-th derivative coefficient.
To perform this expansion, we need the numerical values of $f_1$, $h_1$ as well as $r_h$. 
We will find them by numerically solving the background Eqs.~(\ref{eq:eqphi})-(\ref{dfh}) with the boundary conditions (\ref{fhexpan}) expanded up to sufficiently high orders in $i$.

Far away from the horizon where the metric components are given by $f \simeq h =1-2M/r+\mathcal{O}(r^{-2})$, we have $r_{*} \simeq r+2M \ln\,[r/(2M)-1]$ and hence $e^{i\omega r_{*}} \propto e^{i\omega r}r^{2 i \omega M}$. 
At spatial infinity, this leads to the following ansatz 
\be
\psi_{i}^{\rm I}=e^{i\omega r}r^{2 i\omega M}
\sum_{n=0} (\psi_{i}^{\rm I})^{(n)} r^{-n}\,.
\label{psiI}
\ee
Solving the perturbation equations order by order in the regime 
$r \gg r_h$, we find that the coefficients $(\psi_{i}^{\rm I})^{(n)}$ with $n \geq 1$ are all functions of $(\psi_{i}^{\rm I})^{(0)}$. 
Then the space of independent solutions is five, 
as it is also 
the case for the expansion (\ref{psiH}) around $r=r_h$.

In the following, we will focus on the massless axion ($m_{\phi}=0$) and the quadrupole perturbations ($l=2$).
The computation of QNMs can be easily extended to the massive axion whose Compton wavelength $m_{\phi}^{-1}$ is much larger than $r_h$. 
For the numerical computation, it is useful to perform the rescalings $r=\bar{r}\,r_p$, $q_{M}=\bar{q}_{M} \Mpl r_{p}$, and 
$q_{E}=\bar{q}_{E} \Mpl r_{p}$, where $r_{p}$ is a pivot radius. 
We will first find the value of $r_h$ leading to the BH ADM mass $M=1$ and then choose $r_p=M=1$. 
In the following, we will omit the bars from $\bar{r}$, $\bar{q}_{M}$, and $\bar{q}_{E}$ to keep the notation simpler. 
We also apply this rescaling to the perturbation equations of motion.

%%%%%%%%%%%%%%%%%%%%%%%%%%%%%%%%%%%%%%%%%%%%%%%%%%%%%%%%%%%%%%%%%%%%%
\begin{figure}[hb]
\centering
\includegraphics[height=4.0in,width=3.5in]{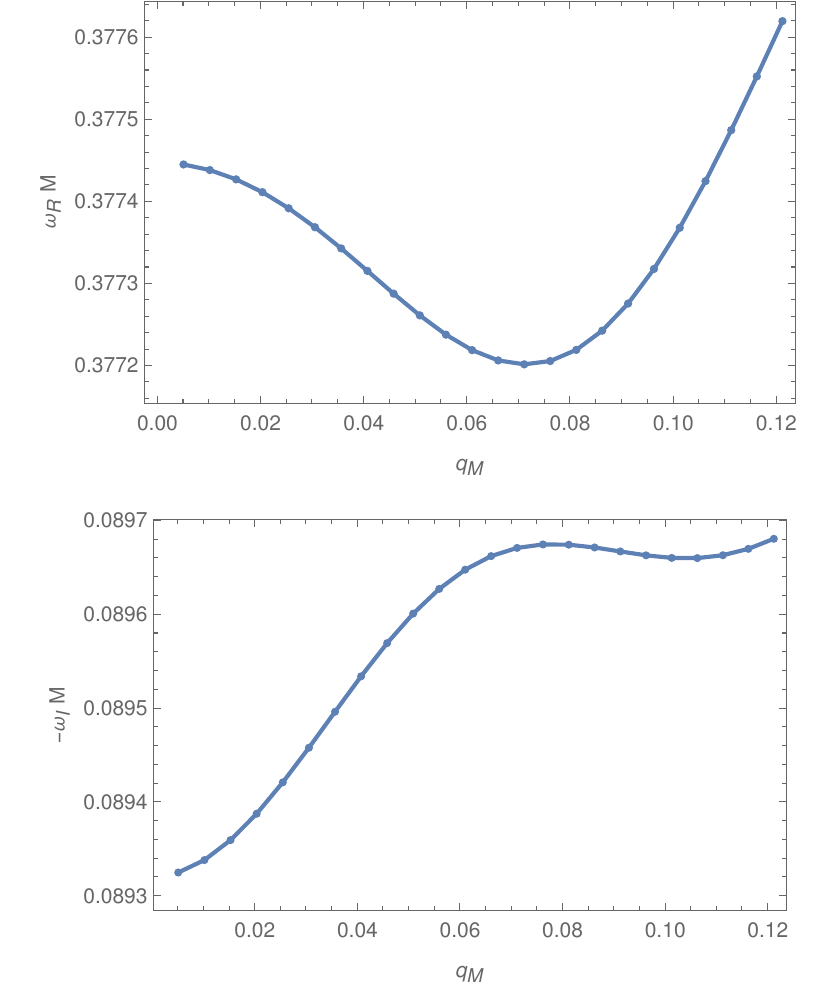}
\caption{Dependence of the $l=2$ fundamental gravitational 
QNM frequencies $\omega=\omega_R+i \omega_I$ 
on the magnetic charge $q_M$. 
The top and bottom panels show $\omega_R M$ 
and $-\omega_I M$ versus $q_M$, respectively.
Each point corresponds to a different value of $q_M$, 
but each point/configuration has the same BH mass 
$M=1$ and total squared BH charge 
$q_M^2+q_E^2=13/50$.}
\label{fig:fund1}
\end{figure}
%%%%%%%%%%%%%%%%%%%%%%%%%%%%%%%%%%%%%%%%%%%%%%%%%%%%%%%%%%%%%%%%%%%%%%

%%%%%%%%%%%%%%%%%%%%%%%%%%%%%%%%%%%%%%%%%%%%%%%%%%%%%%%%%%%%%
\begin{figure}[hb]
\centering
\includegraphics[height=4.0in,width=3.5in]{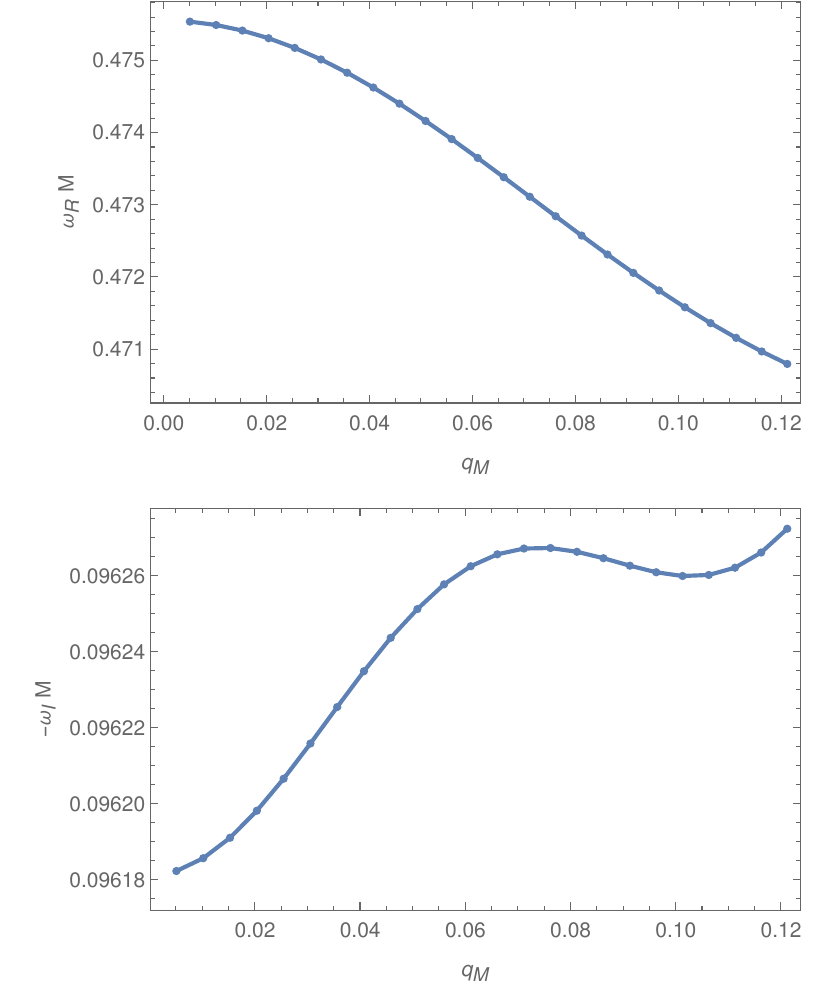}
\caption{Dependence of the $l=2$ fundamental electromagnetic QNM 
frequencies 
on $q_M$. The choices of $q_M$, $q_E$, and $M$ are the same as 
those in Fig.~\ref{fig:fund1}.}
\label{fig:fund2}
\end{figure}
%%%%%%%%%%%%%%%%%%%%%%%%%%%%%%%%%%%%%%%%%%%%%%%%%%%%%%%%%

We vary the values of $q_{E}$ and $q_{M}$ by keeping the BH mass $M$ 
and the total charge $q_T=\sqrt{q_{E}^{2}+q_{M}^{2}}$ constant.
After this, we only have the freedom of choosing five constants 
on the horizon, $(\psi_{i}^{\rm H})^{(0)}$, and the other five 
at infinity, namely $(\psi_{i}^{\rm I})^{(0)}$. 
We can build up ten independent solutions as follows. 
The first solution is found by integrating the perturbation 
equations from the vicinity of $r=r_h$ up to a value of 
$r=r_{{\rm mid}}<\infty$ (typically $r_{{\rm mid}}=5$), 
with the coefficients $(\psi_{1}^{\rm H})^{(0)}=1$ and 
$(\psi_{j}^{\rm H})^{(0)}=0$ for $j\neq1$.
We repeat this procedure by choosing $(\psi_{2}^{\rm H})^{(0)}=1$ and 
$(\psi_{j}^{\rm H})^{(0)}=0$ with $j\neq2$, until we arrive at $i=5$. 
These solutions and radial derivatives, which are evaluated at $r=r_{{\rm mid}}$, are called $\tilde{\psi}_{i,j}^{\rm H}$ and 
${\rm d}\tilde{\psi}_{i,j}^{\rm H}/{\rm d}r$, respectively, where $j$ stands for the nonzero value of $(\psi_{j}^{\rm H})^{(0)}$. 
In this way, we can build a matrix ${\cal A}$ with the first five columns given by $(\tilde{\psi}_{i,j}^{\rm H},{\rm d}\tilde{\psi}_{i,j}^{\rm H}
/{\rm d}r)^{\rm T}$.

We will also find five other independent solutions 
by integrating from 
infinity down to $r=r_{{\rm mid}}$. For the boundary conditions, 
we fix one of the $\psi_{j}^{{\rm I},(0)}$ to 1 and the other 
four elements to zero. 
We call these solutions and radial derivatives 
$\tilde{\psi}_{i,j}^{\rm I}$ and 
${\rm d}\tilde{\psi}_{i,j}^{\rm I}/{\rm d}r$, respectively, 
and again naming by $j$ the nonzero $(\psi_{j}^{\rm I})^{(0)}$. 
Adding the five columns 
$(\tilde{\psi}_{i,j}^{\rm I},{\rm d}\tilde{\psi}_{i,j}^{\rm I}/{\rm d}r)^{\rm T}$ 
to the matrix $\mathcal{A}$, we obtain the $10 \times 10$ matrix $\tilde{{\cal A}}$. 
From the determinant equation $\det \tilde{{\cal A}}=0$, 
we can obtain the QNM frequency $\omega$.

We consider the two fundamental QNMs that are present in the limit $q_M \to 0$.\footnote{If isospectrality is broken, we should have other fundamental frequencies, one from the gravitational side and the other one from the electromagnetic side. In addition, independently of the isospectrality breaking, we should expect to have another frequency coming from the scalar mode, due to its nontrivial hair. This work focuses on the possibility of finding the axion coupled to photons around charged black holes by breaking electric-magnetic duality. 
The detailed study of the whole spectrum of QNM frequencies will be discussed elsewhere.} 
For $q_M=0$, the background solution reduces to the electrically charged RN BH without the axion hair\footnote{In this limit, we can recognize numerically the spectrum of the QNMs as being gravitational or electromagnetic.}. 
In this case, there are one gravitational and the other electromagnetic QNMs, whose frequencies were computed in Refs.~\cite{Kokkotas:1988fm, Gunter:1980, Kokkotas:1988fm}. 
The study of a possible extra fundamental QNM due to the nontrivial axion profile is left for future work. In fact, in this study, we would like to focus on the crucial property of hairy BHs to distinguish the magnetic charge from the electric one at the level of the gravitational/electromagnetic QNMs, leading to an unequivocal sign for the existence of axions. 
We choose a configuration with $q_M=q_T \sin\alpha$ and $q_E=q_T \cos \alpha$, where the total charge $q_T$ is chosen to be $\sqrt{13/50}$.
We vary the angle $\alpha$ from 0 to $12/50$, where each solution differs from the previous one by the interval $\Delta\alpha=1/100$. 
For each value of $\alpha$, we numerically solve the background equations of motion and find the value of $r_h$ leading to $M=1$, so that all the BHs have the same mass and total charge. 

In Fig.~\ref{fig:fund1}, we plot the QNM frequencies $\omega=\omega_R+i \omega_I$ for the gravitational fundamental mode. 
In the limit $q_M \to 0$ we obtain $\omega M=0.37744-0.08932i$, which coincides with the value of an electrically charged RN BH with $q_E=\sqrt{13/50}$. 
For $q_M \neq 0$, both $\omega_R$ and $\omega_I$ change as a function of $q_M$. This property is in stark contrast to the RN BH without the axion-photon coupling, where the QNM is independent of $q_M$ for a fixed total charge $q_T=\sqrt{q_E^2+q_M^2}$ and a mass $M$. The axion-photon coupling breaks this degeneracy of QNMs relevant to electric-magnetic duality \cite{Pereniguez:2023wxf}.

In Fig.~\ref{fig:fund2}, we also show the electromagnetic fundamental frequencies as a function of $q_M$. 
In the limit $q_M \to 0$, we confirm that the electromagnetic QNM approaches the value $\omega=0.4756-0.09618i$ derived for the RN BH with $q_E=\sqrt{13/50}$. 
For $q_M \neq 0$, both the real and imaginary parts of the electromagnetic QNM explicitly depend on the ratio $q_M/q_T$. 
We showed this property for a total charge $q_T=\sqrt{13/50}$, but it also persists for general nonvanishing values of $q_T$. 
Moreover, the overtones of both gravitational and electromagnetic perturbations are also dependent on $q_M/q_T$ 
for fixed values of $q_T$ and $M$. 
Thus, the gravitational-wave observations of QNMs allow us to distinguish between the charged BH with the axion hair and the magnetically (or electrically) charged RN BH.

%=============================================%
\section{IV. Conclusion}
%=============================================%

The magnetically charged BH can be present today as a remnant 
of the absorption of magnetic monopoles in the early Universe. 
For a given total charge $q_T$ and mass $M$, the QNMs of RN BHs are the same independent of the mixture of magnetic and electric charges. 
In the presence of the axion coupled to photons, however, we showed 
that the charged BH with the axion hair breaks this degeneracy. 
We computed the gravitational and electromagnetic QNMs and found that both QNMs depend on the ratio $q_M/q_T$ for hairy BH solutions realized by the axion-photon coupling.
Hence the upcoming high-precision observations of QNMs offer the possibility for detecting the signatures of both the magnetic charge and the axion.

There are several interesting extensions of our work. 
First, the computation of QNMs for charged rotating BH solutions with the axion hair \cite{Burrage:2023zvk} is the important next step for placing realistic bounds on our model parameters. 
Next, the gravitational waveforms emitted during the inspiral phase of charged binary BHs with the axion hair will put further constraints on the theory. 
Thirdly, the observations of BH shadows such as the Event Horizon Telescope \cite{EventHorizonTelescope:2021dqv} will give upper bounds on the BH charges. 
Fourth, we leave a detailed study of the isospectrality of QNMs for a future separate 
work. While isospectrality may be broken, this letter aims to demonstrate the potential for simultaneously finding magnetic charges and axions through 
the QNMs of BHs. 
Finally, it will be of interest to study the effect of large magnetic fields on the BH physics near the horizon, e.g., restoration of an electroweak symmetry \cite{Maldacena:2020skw}. 
These issues are left for future work.

%=============================================%
\acknowledgments
\section{Acknowledgement}
The work of ADF was supported by the Japan Society for the Promotion of Science Grants-in-Aid for Scientific Research No.\ 20K03969. 
ST was supported by the Grant-in-Aid for Scientific Research 
Fund of the JSPS No.~22K03642 and Waseda University Special Research Project No.~2023C-473. 
%=============================================%

\appendix

\section{Appendix A:~Second-order action of perturbations}
\label{AppenA}

After integrating the second-order action of perturbations 
with respect to $\theta$ and $\varphi$ and performing integration 
by parts, the resulting quadratic-order action  
can be expressed in the form 
${\cal S}^{(2)}=\int \rd t \rd r\,{\cal L}$, 
with the Lagrangian

\begin{widetext}
\ba
{\cal L} &=&
p_1 \left( \dot{W}-Q'+\frac{2Q}{r} \right)^2
+p_2  \left( \dot{W}-Q'+\frac{2Q}{r} \right) \delta A
+p_3 \left( Q^2-fh W^2 \right)
+p_4 \left( \dot{\delta A}^2-fh \delta A'^2
-\frac{fL}{r^2} \delta A^2 \right) \nonumber \\
& &+p_5 Q \delta A+p_6 Q h_1
+p_7 \left( Q \delta A_0-fh W \delta A_1 \right)
+p_8[2\gc (\gc q_M \phi + q_E) \delta A \delta \phi  
+ q_M \{ 2h \delta A'\,h_1 - (H_0 - H_2) \delta A \}]\nonumber \\
& &+\gc [ L \phi' \delta A \delta A_0
-q_M \delta \phi (\delta A_0' - \dot{\delta A}_1)]\nonumber \\
& &+
a_0 H_0^2 
+ H_0 \left[ a_1 \delta \phi' + a_2 H_2' + L a_3 h_1' 
+a_4 \delta \phi + (a_5+L a_6)H_2 + L a_7 h_1 \right]
+Lb_0 H_1^2+H_1 (b_1 \dot{\delta \phi}
+b_2 \dot{H}_2+L b_3 \dot{h}_1)\nonumber \\
& &+c_0 H_2^2 + H_2 ( c_1 \delta \phi' 
+ c_2 \delta \phi+L c_3 h_1)+
L d_1 \dot{h}_1^2+L d_2 h_1 \delta \phi+L d_3 h_1^2
+e_1 \dot{\delta \phi}^2 + e_2 \delta \phi'^2 
+ (e_3+L e_4 ) \delta \phi^2\nonumber \\
& & +s_1 ( \delta A_0'-\dot{\delta A_1} )^2
+s_2 (H_0 - H_2)( \delta A_0'-\dot{\delta A_1} )
+L ( s_3 h_1 \delta A_0 + s_4 \delta A_0^2 
+ s_5 \delta A_1^2)\,,
\label{Lag}
\ea
where $L \equiv l(l+1)$, and 
\ba
& &
p_1=\frac{L\Mpl^2 \sqrt{h}}{4\sqrt{f}},\quad 
p_2=-\frac{L(\gc q_M \phi+q_E)}{r^2},\quad
p_3=\frac{L (L \Mpl^2 r^2 -2 \Mpl^2 r^2 + 2 q_M^2)}
{4r^4 \sqrt{fh}},\quad 
p_4=\frac{L}{2\sqrt{fh}},\quad 
p_5=\frac{L \gc q_M \phi'}{r^2},\nonumber \\
& &
p_6=\frac{q_M}{r^2}p_2,\quad 
p_7=\frac{L q_M}{r^2 \sqrt{fh}},\quad
p_8=-\frac{L\sqrt{f}}{2r^2 \sqrt{h}},\quad
a_0=\frac{\sqrt{f}(\gc q_M \phi+q_E)^2}{8r^2 \sqrt{h}},
\quad
a_1=\frac{r^2 \phi' \sqrt{fh}}{2},\quad 
a_2=-\frac{r \Mpl^2 \sqrt{fh}}{2},\nonumber \\
& &
a_3=\frac{\Mpl^2 \sqrt{fh}}{2},\quad
a_4=\frac{r^2 \sqrt{f}\,m_{\phi}^2 \phi}{2\sqrt{h}},\quad
a_5=\frac{\sqrt{f}(m_{\phi}^2 \phi^2 r^4-2\Mpl^2 r^2+q_M^2)}
{4r^2 \sqrt{h}},\quad 
a_6=-\frac{\Mpl^2 \sqrt{f}}{4 \sqrt{h}},\nonumber \\
& &
a_7=-\frac{\sqrt{f}\,[(h \phi'^2+m_{\phi}^2 \phi^2)r^4-2\Mpl^2 r^2 (h+1)
+q_M^2+(\gc q_M \phi+q_E)^2]}{8 r^3 \sqrt{h}},\quad 
b_0=\frac{\Mpl^2 \sqrt{h}}{4\sqrt{f}},\quad
b_1=-\frac{r^2 \phi'\sqrt{h}}{\sqrt{f}},\nonumber \\
& &
b_2=4b_0 r,\quad
b_3=-2b_0,\quad
c_0=-\frac{a_5}{2},\quad 
c_1=a_1,\quad
c_2=-a_4,\quad 
c_3=-a_7-\frac{r \phi'^2 \sqrt{fh}}{4},\quad
d_1=b_0,\quad d_2=\frac{2}{r^2}a_1,\nonumber \\
\hspace{-1.0cm}
& &
d_3=\frac{\sqrt{fh}(\Mpl^2 r^2-q_M^2)}{2r^4},\quad
e_1=\frac{r^2}{2\sqrt{fh}},\quad 
e_2=-\frac{r^2 \sqrt{fh}}{2},\quad
e_3=-\frac{r^2 \sqrt{f}\,m_{\phi}^2}{2\sqrt{h}},\quad
e_4=-\frac{\sqrt{f}}{2\sqrt{h}},\nonumber \\
\hspace{-1.0cm}
& &
s_1=\frac{r^2}{2}\sqrt{\frac{h}{f}},\quad
s_2=\frac{\gc q_M \phi+q_E}{2},\quad 
s_3=-\frac{2}{r^2}s_2,\quad
s_4=\frac{e_1}{r^2},\quad 
s_5=-\frac{a_3}{\Mpl^2}\,.
\ea
\end{widetext}
Note that a similar second-order action of odd- and even-parity 
perturbations in Maxwell-Horndeski theories with $q_M=0$ 
and $q_E \neq 0$ was derived in Ref.~\cite{Kase:2023kvq}. 
In current EMA theory, the existence of the nonvanishing 
magnetic charge $q_M$
does not allow the separation of ${\cal S}^{(2)}$ into the 
odd- and even-parity modes.

\section{Appendix B:~Dynamical perturbations}
\label{AppenB}

Since some of the perturbed variables appearing in the 
Lagrangian (\ref{Lag}) are nondynamical, they can be integrated out 
from the second-order action. 
For the fields associated with $Q$ and $W$, we introduce a Lagrangian 
multiplier $\chi_1$ as 
\begin{equation}
{\cal L}_2 = {\cal L} + {\bar b}_1\,\left(\dot W - Q' +\frac{2Q}{r} 
+ {\bar b}_2 \delta A -\chi_1 \right)^2\,,
\label{L2}
\end{equation}
where a dot represents the derivative with respect to $t$.
The coefficients ${\bar b}_1$ and ${\bar b}_2$ are chosen to 
remove the products $\dot{W}^2$, $Q'^2$, and 
$\delta A\, Q'$ from ${\cal L}_2$.
Then, we find
\begin{equation}
{\bar b}_1 = -\frac{\Mpl^2 l(l+1)\sqrt{h}}{4\sqrt{f}}\,,\quad
{\bar b}_2 = -\frac{2\sqrt{f}(\gc \phi q_M + q_E)}
{\Mpl^2 r^2\sqrt{h}}\,.
\end{equation}
At this point, both $Q$ and $W$ can be eliminated from 
the action by employing their equations of motion. 
Varying ${\cal L}_2$ with respect to $\chi_1$, we obtain 
\begin{equation}
\chi_1=\dot W - Q' +\frac{2Q}{r} 
-\frac{2\sqrt{f}(\gc \phi q_M + q_E)}
{\Mpl^2 r^2\sqrt{h}} \delta A\,,
\label{chi1}
\end{equation}
with which ${\cal L}_2$ is equivalent to ${\cal L}$.

After several integrations by parts, one can remove the 
nondynamical perturbation $H_{1}$ from ${\cal L}_2$  
by using its equation of motion. 
After this process, we introduce a new field 
\begin{equation}
v_{1}=H_{2}-\frac{l(l+1)}{r} h_{1}
-\frac{r\phi'\delta\phi}{\Mpl^2}\,,
\label{v1}
\end{equation}
together with the other Lagrange multiplier 
$\chi_{2}$, as
\ba
{\cal L}_{3} &=& {\cal L}_{2}-\frac{r^{2}\sqrt{h}}{2\sqrt{f}} 
[\dot{\delta A}_{1}-\delta A'_{0}
-\bar{c}_1 H_{0}+\bar{c}_2 h_{1} 
\nonumber \\
& &\qquad \qquad \quad 
\,+\bar{c}_3 v_{1}+\bar{c}_4 \delta\phi-\chi_{2} ]^{2}\,.
\ea
The coefficients $\bar{c}_i$ (where $i=1,2,3,4$) are chosen to 
obtain the reduced Lagrangian for the propagating degrees of 
freedom with a reasonably simple form. 
On choosing
\begin{equation}
\bar{c}_1=\frac{\sqrt{f}}{2r^{2}\sqrt{h}}\,
(q_{E}+\gc q_{M}\phi)\,,
\end{equation}
the terms $\dot{\delta A}_1^2$, $\delta A_0'^2$, $\dot{\delta A}_1 \delta A_0'$, $H_{0}^{2}$, and $H_{0}\dot{\delta A}_{1}$ are vanishing.
Furthermore, we set $\bar{c}_2=l(l+1)\bar{c}_1/r$ 
and $\bar{c}_3=\bar{c}_1$ to 
eliminate the products $h_{1}\dot{\delta A}_{1}$ and 
$v_{1}\dot{\delta A}_{1}$, respectively.
Finally, we choose 
\begin{equation}
\bar{c}_4=\frac{\sqrt{f}}{2r^{2}\sqrt{h}}\,
\left[ 2\gc q_{M}+\frac{r\phi'}{\Mpl^2}
(q_{E}+\gc q_{M}\phi) \right]\,,
\end{equation}
to remove the term $\delta\phi\,\dot{\delta A}_{1}$.
At this point, $H_{0}$ becomes a Lagrangian multiplier and its equation of motion sets a constraint for other perturbations. 
This equation can be solved algebraically for $h_{1}$. Varying ${\cal L}_3$ with respect to $\chi_2$, we obtain
\ba
\hspace{-0.65cm}
& &
\chi_2= 
\dot{\delta A}_{1}-\delta A'_{0} \nonumber \\
\hspace{-0.65cm}
& & \qquad
+\frac{\sqrt{f}}{2r^{2}\sqrt{h}}\,
(q_{E}+\gc q_{M}\phi) \left[ v_1+\frac{l(l+1)}{r}h_1-H_0 
\right] \nonumber \\
\hspace{-0.65cm}
& & \qquad
+\frac{\sqrt{f}}{2r^{2}\sqrt{h}}\,
\left[ 2\gc q_{M}+\frac{r\phi'}{\Mpl^2}
(q_{E}+\gc q_{M}\phi) \right] \delta \phi.
\label{chi2}
\ea
The introduction of $\chi_{2}$ makes both $\delta A_{0}$ and $\delta A_{1}$ Lagrange multipliers, so that they can be 
removed from the action.

After this procedure, the resulting second-order action contains 
only five dynamical perturbations: $\chi_{1}$,
$v_{1}$, $\chi_{2}$, $\delta A$, $\delta\phi$, and 
$t, r$ derivatives.
For high radial and angular momentum modes, we can show that 
the ghosts are absent  
and all the dynamical fields propagate with the speed of light.

\bibliographystyle{mybibstyle}
\bibliography{bib}

%merlin.mbs apsrev4-1.bst 2010-07-25 4.21a (PWD, AO, DPC) hacked
%Control: key (0)
%Control: author (72) initials jnrlst
%Control: editor formatted (1) identically to author
%Control: production of article title (-1) disabled
%Control: page (0) single
%Control: year (1) truncated
%Control: production of eprint (0) enabled
\begin{thebibliography}{40}%
\makeatletter
\providecommand \@ifxundefined [1]{%
 \@ifx{#1\undefined}
}%
\providecommand \@ifnum [1]{%
 \ifnum #1\expandafter \@firstoftwo
 \else \expandafter \@secondoftwo
 \fi
}%
\providecommand \@ifx [1]{%
 \ifx #1\expandafter \@firstoftwo
 \else \expandafter \@secondoftwo
 \fi
}%
\providecommand \natexlab [1]{#1}%
\providecommand \enquote  [1]{``#1''}%
\providecommand \bibnamefont  [1]{#1}%
\providecommand \bibfnamefont [1]{#1}%
\providecommand \citenamefont [1]{#1}%
\providecommand \href@noop [0]{\@secondoftwo}%
\providecommand \href [0]{\begingroup \@sanitize@url \@href}%
\providecommand \@href[1]{\@@startlink{#1}\@@href}%
\providecommand \@@href[1]{\endgroup#1\@@endlink}%
\providecommand \@sanitize@url [0]{\catcode `\\12\catcode `\$12\catcode
  `\&12\catcode `\#12\catcode `\^12\catcode `\_12\catcode `\%12\relax}%
\providecommand \@@startlink[1]{}%
\providecommand \@@endlink[0]{}%
\providecommand \url  [0]{\begingroup\@sanitize@url \@url }%
\providecommand \@url [1]{\endgroup\@href {#1}{\urlprefix }}%
\providecommand \urlprefix  [0]{URL }%
\providecommand \Eprint [0]{\href }%
\providecommand \doibase [0]{http://dx.doi.org/}%
\providecommand \selectlanguage [0]{\@gobble}%
\providecommand \bibinfo  [0]{\@secondoftwo}%
\providecommand \bibfield  [0]{\@secondoftwo}%
\providecommand \translation [1]{[#1]}%
\providecommand \BibitemOpen [0]{}%
\providecommand \bibitemStop [0]{}%
\providecommand \bibitemNoStop [0]{.\EOS\space}%
\providecommand \EOS [0]{\spacefactor3000\relax}%
\providecommand \BibitemShut  [1]{\csname bibitem#1\endcsname}%
\let\auto@bib@innerbib\@empty
%</preamble>
\bibitem [{\citenamefont {Abbott}\ \emph {et~al.}(2016)\citenamefont {Abbott}
  \emph {et~al.}}]{LIGOScientific:2016aoc}%
  \BibitemOpen
  \bibfield  {author} {\bibinfo {author} {\bibfnamefont {B.~P.}\ \bibnamefont
  {Abbott}} \emph {et~al.} (\bibinfo {collaboration} {LIGO Scientific,
  Virgo}),\ }\href {\doibase 10.1103/PhysRevLett.116.061102} {\bibfield
  {journal} {\bibinfo  {journal} {\emph {Phys. Rev. Lett.}}\ }\textbf {\bibinfo
  {volume} {116}},\ \bibinfo {pages} {061102} (\bibinfo {year} {2016})},\
  \Eprint {http://arxiv.org/abs/1602.03837} {arXiv:1602.03837 [gr-qc]}
  \BibitemShut {NoStop}%
\bibitem [{\citenamefont {Kokkotas}\ and\ \citenamefont
  {Schmidt}(1999)}]{Kokkotas:1999bd}%
  \BibitemOpen
  \bibfield  {author} {\bibinfo {author} {\bibfnamefont {K.~D.}\ \bibnamefont
  {Kokkotas}} and \bibinfo {author} {\bibfnamefont {B.~G.}\ \bibnamefont
  {Schmidt}},\ }\href {\doibase 10.12942/lrr-1999-2} {\bibfield  {journal}
  {\bibinfo  {journal} {\emph {Living Rev. Rel.}}\ }\textbf {\bibinfo {volume}
  {2}},\ \bibinfo {pages} {2} (\bibinfo {year} {1999})},\ \Eprint
  {http://arxiv.org/abs/gr-qc/9909058} {arXiv:gr-qc/9909058} \BibitemShut
  {NoStop}%
\bibitem [{\citenamefont {Nollert}(1999)}]{Nollert:1999ji}%
  \BibitemOpen
  \bibfield  {author} {\bibinfo {author} {\bibfnamefont {H.-P.}\ \bibnamefont
  {Nollert}},\ }\href {\doibase 10.1088/0264-9381/16/12/201} {\bibfield
  {journal} {\bibinfo  {journal} {\emph {Class. Quant. Grav.}}\ }\textbf
  {\bibinfo {volume} {16}},\ \bibinfo {pages} {R159} (\bibinfo {year}
  {1999})}\BibitemShut {NoStop}%
\bibitem [{\citenamefont {Berti}\ \emph {et~al.}(2009)\citenamefont {Berti},
  \citenamefont {Cardoso},\ and\ \citenamefont {Starinets}}]{Berti:2009kk}%
  \BibitemOpen
  \bibfield  {author} {\bibinfo {author} {\bibfnamefont {E.}~\bibnamefont
  {Berti}}, \bibinfo {author} {\bibfnamefont {V.}~\bibnamefont {Cardoso}},  and
  \bibinfo {author} {\bibfnamefont {A.~O.}\ \bibnamefont {Starinets}},\ }\href
  {\doibase 10.1088/0264-9381/26/16/163001} {\bibfield  {journal} {\bibinfo
  {journal} {\emph {Class. Quant. Grav.}}\ }\textbf {\bibinfo {volume} {26}},\
  \bibinfo {pages} {163001} (\bibinfo {year} {2009})},\ \Eprint
  {http://arxiv.org/abs/0905.2975} {arXiv:0905.2975 [gr-qc]} \BibitemShut
  {NoStop}%
\bibitem [{\citenamefont {Konoplya}\ and\ \citenamefont
  {Zhidenko}(2011)}]{Konoplya:2011qq}%
  \BibitemOpen
  \bibfield  {author} {\bibinfo {author} {\bibfnamefont {R.~A.}\ \bibnamefont
  {Konoplya}} and \bibinfo {author} {\bibfnamefont {A.}~\bibnamefont
  {Zhidenko}},\ }\href {\doibase 10.1103/RevModPhys.83.793} {\bibfield
  {journal} {\bibinfo  {journal} {\emph {Rev. Mod. Phys.}}\ }\textbf {\bibinfo
  {volume} {83}},\ \bibinfo {pages} {793} (\bibinfo {year} {2011})},\ \Eprint
  {http://arxiv.org/abs/1102.4014} {arXiv:1102.4014 [gr-qc]} \BibitemShut
  {NoStop}%
\bibitem [{\citenamefont {Pani}(2013)}]{Pani:2013pma}%
  \BibitemOpen
  \bibfield  {author} {\bibinfo {author} {\bibfnamefont {P.}~\bibnamefont
  {Pani}},\ }\href {\doibase 10.1142/S0217751X13400186} {\bibfield  {journal}
  {\bibinfo  {journal} {\emph {Int. J. Mod. Phys. A}}\ }\textbf {\bibinfo
  {volume} {28}},\ \bibinfo {pages} {1340018} (\bibinfo {year} {2013})},\
  \Eprint {http://arxiv.org/abs/1305.6759} {arXiv:1305.6759 [gr-qc]}
  \BibitemShut {NoStop}%
\bibitem [{\citenamefont {Guo}\ \emph {et~al.}(2023)\citenamefont {Guo},
  \citenamefont {Tan},\ and\ \citenamefont {Liu}}]{Guo:2022rms}%
  \BibitemOpen
  \bibfield  {author} {\bibinfo {author} {\bibfnamefont {W.-D.}\ \bibnamefont
  {Guo}}, \bibinfo {author} {\bibfnamefont {Q.}~\bibnamefont {Tan}},  and
  \bibinfo {author} {\bibfnamefont {Y.-X.}\ \bibnamefont {Liu}},\ }\href
  {\doibase 10.1103/PhysRevD.107.124046} {\bibfield  {journal} {\bibinfo
  {journal} {\emph {Phys. Rev. D}}\ }\textbf {\bibinfo {volume} {107}},\
  \bibinfo {pages} {124046} (\bibinfo {year} {2023})},\ \Eprint
  {http://arxiv.org/abs/2212.08784} {arXiv:2212.08784 [gr-qc]} \BibitemShut
  {NoStop}%
\bibitem [{\citenamefont {Gunter}(1980)}]{Gunter:1980}%
  \BibitemOpen
  \bibfield  {author} {\bibinfo {author} {\bibfnamefont {D.~L.}\ \bibnamefont
  {Gunter}},\ }\href {\doibase 10.1098/rsta.1980.0190} {\bibfield  {journal}
  {\bibinfo  {journal} {\emph {Phil. Trans. Roy. Soc. Lond}}\ }\textbf
  {\bibinfo {volume} {A296}},\ \bibinfo {pages} {497} (\bibinfo {year}
  {1980})}\BibitemShut {NoStop}%
\bibitem [{\citenamefont {Kokkotas}\ and\ \citenamefont
  {Schutz}(1988)}]{Kokkotas:1988fm}%
  \BibitemOpen
  \bibfield  {author} {\bibinfo {author} {\bibfnamefont {K.~D.}\ \bibnamefont
  {Kokkotas}} and \bibinfo {author} {\bibfnamefont {B.~F.}\ \bibnamefont
  {Schutz}},\ }\href {\doibase 10.1103/PhysRevD.37.3378} {\bibfield  {journal}
  {\bibinfo  {journal} {\emph {Phys. Rev. D}}\ }\textbf {\bibinfo {volume}
  {37}},\ \bibinfo {pages} {3378} (\bibinfo {year} {1988})}\BibitemShut
  {NoStop}%
\bibitem [{\citenamefont {Leaver}(1990)}]{Leaver:1990zz}%
  \BibitemOpen
  \bibfield  {author} {\bibinfo {author} {\bibfnamefont {E.~W.}\ \bibnamefont
  {Leaver}},\ }\href {\doibase 10.1103/PhysRevD.41.2986} {\bibfield  {journal}
  {\bibinfo  {journal} {\emph {Phys. Rev. D}}\ }\textbf {\bibinfo {volume}
  {41}},\ \bibinfo {pages} {2986} (\bibinfo {year} {1990})}\BibitemShut
  {NoStop}%
\bibitem [{\citenamefont {Berti}\ and\ \citenamefont
  {Kokkotas}(2003)}]{Berti:2003zu}%
  \BibitemOpen
  \bibfield  {author} {\bibinfo {author} {\bibfnamefont {E.}~\bibnamefont
  {Berti}} and \bibinfo {author} {\bibfnamefont {K.~D.}\ \bibnamefont
  {Kokkotas}},\ }\href {\doibase 10.1103/PhysRevD.68.044027} {\bibfield
  {journal} {\bibinfo  {journal} {\emph {Phys. Rev. D}}\ }\textbf {\bibinfo
  {volume} {68}},\ \bibinfo {pages} {044027} (\bibinfo {year} {2003})},\
  \Eprint {http://arxiv.org/abs/hep-th/0303029} {arXiv:hep-th/0303029}
  \BibitemShut {NoStop}%
\bibitem [{\citenamefont {Maldacena}(2021)}]{Maldacena:2020skw}%
  \BibitemOpen
  \bibfield  {author} {\bibinfo {author} {\bibfnamefont {J.}~\bibnamefont
  {Maldacena}},\ }\href {\doibase 10.1007/JHEP04(2021)079} {\bibfield
  {journal} {\bibinfo  {journal} {\emph {JHEP}}\ }\textbf {\bibinfo {volume}
  {04}},\ \bibinfo {pages} {079} (\bibinfo {year} {2021})},\ \Eprint
  {http://arxiv.org/abs/2004.06084} {arXiv:2004.06084 [hep-th]} \BibitemShut
  {NoStop}%
\bibitem [{\citenamefont {Stojkovic}\ and\ \citenamefont
  {Freese}(2005)}]{Stojkovic:2004hz}%
  \BibitemOpen
  \bibfield  {author} {\bibinfo {author} {\bibfnamefont {D.}~\bibnamefont
  {Stojkovic}} and \bibinfo {author} {\bibfnamefont {K.}~\bibnamefont
  {Freese}},\ }\href {\doibase 10.1016/j.physletb.2004.12.019} {\bibfield
  {journal} {\bibinfo  {journal} {\emph {Phys. Lett. B}}\ }\textbf {\bibinfo
  {volume} {606}},\ \bibinfo {pages} {251} (\bibinfo {year} {2005})},\ \Eprint
  {http://arxiv.org/abs/hep-ph/0403248} {arXiv:hep-ph/0403248} \BibitemShut
  {NoStop}%
\bibitem [{\citenamefont {Kobayashi}(2021)}]{Kobayashi:2021des}%
  \BibitemOpen
  \bibfield  {author} {\bibinfo {author} {\bibfnamefont {T.}~\bibnamefont
  {Kobayashi}},\ }\href {\doibase 10.1103/PhysRevD.104.043501} {\bibfield
  {journal} {\bibinfo  {journal} {\emph {Phys. Rev. D}}\ }\textbf {\bibinfo
  {volume} {104}},\ \bibinfo {pages} {043501} (\bibinfo {year} {2021})},\
  \Eprint {http://arxiv.org/abs/2105.12776} {arXiv:2105.12776 [hep-ph]}
  \BibitemShut {NoStop}%
\bibitem [{\citenamefont {Das}\ and\ \citenamefont {Hook}(2021)}]{Das:2021wei}%
  \BibitemOpen
  \bibfield  {author} {\bibinfo {author} {\bibfnamefont {S.}~\bibnamefont
  {Das}} and \bibinfo {author} {\bibfnamefont {A.}~\bibnamefont {Hook}},\
  }\href {\doibase 10.1007/JHEP12(2021)145} {\bibfield  {journal} {\bibinfo
  {journal} {\emph {JHEP}}\ }\textbf {\bibinfo {volume} {12}},\ \bibinfo
  {pages} {145} (\bibinfo {year} {2021})},\ \Eprint
  {http://arxiv.org/abs/2109.00039} {arXiv:2109.00039 [hep-ph]} \BibitemShut
  {NoStop}%
\bibitem [{\citenamefont {Estes}\ \emph {et~al.}(2023)\citenamefont {Estes},
  \citenamefont {Kavic}, \citenamefont {Liebling}, \citenamefont {Lippert},\
  and\ \citenamefont {Simonetti}}]{Estes:2022buj}%
  \BibitemOpen
  \bibfield  {author} {\bibinfo {author} {\bibfnamefont {J.}~\bibnamefont
  {Estes}}, \bibinfo {author} {\bibfnamefont {M.}~\bibnamefont {Kavic}},
  \bibinfo {author} {\bibfnamefont {S.~L.}\ \bibnamefont {Liebling}}, \bibinfo
  {author} {\bibfnamefont {M.}~\bibnamefont {Lippert}},  and \bibinfo {author}
  {\bibfnamefont {J.~H.}\ \bibnamefont {Simonetti}},\ }\href {\doibase
  10.1088/1475-7516/2023/06/017} {\bibfield  {journal} {\bibinfo  {journal}
  {\emph {JCAP}}\ }\textbf {\bibinfo {volume} {06}},\ \bibinfo {pages} {017}
  (\bibinfo {year} {2023})},\ \Eprint {http://arxiv.org/abs/2209.06060}
  {arXiv:2209.06060 [astro-ph.HE]} \BibitemShut {NoStop}%
\bibitem [{\citenamefont {Zhang}\ and\ \citenamefont
  {Zhang}(2023)}]{Zhang:2023tfv}%
  \BibitemOpen
  \bibfield  {author} {\bibinfo {author} {\bibfnamefont {C.}~\bibnamefont
  {Zhang}} and \bibinfo {author} {\bibfnamefont {X.}~\bibnamefont {Zhang}},\
  }\href {\doibase 10.1007/JHEP10(2023)037} {\bibfield  {journal} {\bibinfo
  {journal} {\emph {JHEP}}\ }\textbf {\bibinfo {volume} {10}},\ \bibinfo
  {pages} {037} (\bibinfo {year} {2023})},\ \Eprint
  {http://arxiv.org/abs/2302.07002} {arXiv:2302.07002 [hep-ph]} \BibitemShut
  {NoStop}%
\bibitem [{\citenamefont {Bai}\ and\ \citenamefont
  {Korwar}(2021)}]{Bai:2020ezy}%
  \BibitemOpen
  \bibfield  {author} {\bibinfo {author} {\bibfnamefont {Y.}~\bibnamefont
  {Bai}} and \bibinfo {author} {\bibfnamefont {M.}~\bibnamefont {Korwar}},\
  }\href {\doibase 10.1007/JHEP04(2021)119} {\bibfield  {journal} {\bibinfo
  {journal} {\emph {JHEP}}\ }\textbf {\bibinfo {volume} {04}},\ \bibinfo
  {pages} {119} (\bibinfo {year} {2021})},\ \Eprint
  {http://arxiv.org/abs/2012.15430} {arXiv:2012.15430 [hep-ph]} \BibitemShut
  {NoStop}%
\bibitem [{\citenamefont {De~Felice}\ and\ \citenamefont
  {Tsujikawa}(2024)}]{DeFelice:2023rra}%
  \BibitemOpen
  \bibfield  {author} {\bibinfo {author} {\bibfnamefont {A.}~\bibnamefont
  {De~Felice}} and \bibinfo {author} {\bibfnamefont {S.}~\bibnamefont
  {Tsujikawa}},\ }\href {\doibase 10.1103/PhysRevD.109.084022} {\bibfield
  {journal} {\bibinfo  {journal} {\emph {Phys. Rev. D}}\ }\textbf {\bibinfo
  {volume} {109}},\ \bibinfo {pages} {084022} (\bibinfo {year} {2024})},\
  \Eprint {http://arxiv.org/abs/2312.03191} {arXiv:2312.03191 [gr-qc]}
  \BibitemShut {NoStop}%
\bibitem [{\citenamefont {Nomura}\ \emph {et~al.}(2020)\citenamefont {Nomura},
  \citenamefont {Yoshida},\ and\ \citenamefont {Soda}}]{Nomura:2020tpc}%
  \BibitemOpen
  \bibfield  {author} {\bibinfo {author} {\bibfnamefont {K.}~\bibnamefont
  {Nomura}}, \bibinfo {author} {\bibfnamefont {D.}~\bibnamefont {Yoshida}},
  and \bibinfo {author} {\bibfnamefont {J.}~\bibnamefont {Soda}},\ }\href
  {\doibase 10.1103/PhysRevD.101.124026} {\bibfield  {journal} {\bibinfo
  {journal} {\emph {Phys. Rev. D}}\ }\textbf {\bibinfo {volume} {101}},\
  \bibinfo {pages} {124026} (\bibinfo {year} {2020})},\ \Eprint
  {http://arxiv.org/abs/2004.07560} {arXiv:2004.07560 [gr-qc]} \BibitemShut
  {NoStop}%
\bibitem [{\citenamefont {Nomura}\ and\ \citenamefont
  {Yoshida}(2022)}]{Nomura:2021efi}%
  \BibitemOpen
  \bibfield  {author} {\bibinfo {author} {\bibfnamefont {K.}~\bibnamefont
  {Nomura}} and \bibinfo {author} {\bibfnamefont {D.}~\bibnamefont {Yoshida}},\
  }\href {\doibase 10.1103/PhysRevD.105.044006} {\bibfield  {journal} {\bibinfo
   {journal} {\emph {Phys. Rev. D}}\ }\textbf {\bibinfo {volume} {105}},\
  \bibinfo {pages} {044006} (\bibinfo {year} {2022})},\ \Eprint
  {http://arxiv.org/abs/2111.06273} {arXiv:2111.06273 [gr-qc]} \BibitemShut
  {NoStop}%
\bibitem [{\citenamefont {Pere\~niguez}(2023)}]{Pereniguez:2023wxf}%
  \BibitemOpen
  \bibfield  {author} {\bibinfo {author} {\bibfnamefont {D.}~\bibnamefont
  {Pere\~niguez}},\ }\href {\doibase 10.1103/PhysRevD.108.084046} {\bibfield
  {journal} {\bibinfo  {journal} {\emph {Phys. Rev. D}}\ }\textbf {\bibinfo
  {volume} {108}},\ \bibinfo {pages} {084046} (\bibinfo {year} {2023})},\
  \Eprint {http://arxiv.org/abs/2302.10942} {arXiv:2302.10942 [gr-qc]}
  \BibitemShut {NoStop}%
\bibitem [{\citenamefont {Peccei}\ and\ \citenamefont
  {Quinn}(1977)}]{Peccei:1977hh}%
  \BibitemOpen
  \bibfield  {author} {\bibinfo {author} {\bibfnamefont {R.~D.}\ \bibnamefont
  {Peccei}} and \bibinfo {author} {\bibfnamefont {H.~R.}\ \bibnamefont
  {Quinn}},\ }\href {\doibase 10.1103/PhysRevLett.38.1440} {\bibfield
  {journal} {\bibinfo  {journal} {\emph {Phys. Rev. Lett.}}\ }\textbf {\bibinfo
  {volume} {38}},\ \bibinfo {pages} {1440} (\bibinfo {year}
  {1977})}\BibitemShut {NoStop}%
\bibitem [{\citenamefont {Svrcek}\ and\ \citenamefont
  {Witten}(2006)}]{Svrcek:2006yi}%
  \BibitemOpen
  \bibfield  {author} {\bibinfo {author} {\bibfnamefont {P.}~\bibnamefont
  {Svrcek}} and \bibinfo {author} {\bibfnamefont {E.}~\bibnamefont {Witten}},\
  }\href {\doibase 10.1088/1126-6708/2006/06/051} {\bibfield  {journal}
  {\bibinfo  {journal} {\emph {JHEP}}\ }\textbf {\bibinfo {volume} {06}},\
  \bibinfo {pages} {051} (\bibinfo {year} {2006})},\ \Eprint
  {http://arxiv.org/abs/hep-th/0605206} {arXiv:hep-th/0605206} \BibitemShut
  {NoStop}%
\bibitem [{\citenamefont {Lee}\ and\ \citenamefont
  {Weinberg}(1991)}]{Lee:1991jw}%
  \BibitemOpen
  \bibfield  {author} {\bibinfo {author} {\bibfnamefont {K.-M.}\ \bibnamefont
  {Lee}} and \bibinfo {author} {\bibfnamefont {E.~J.}\ \bibnamefont
  {Weinberg}},\ }\href {\doibase 10.1103/PhysRevD.44.3159} {\bibfield
  {journal} {\bibinfo  {journal} {\emph {Phys. Rev. D}}\ }\textbf {\bibinfo
  {volume} {44}},\ \bibinfo {pages} {3159} (\bibinfo {year}
  {1991})}\BibitemShut {NoStop}%
\bibitem [{\citenamefont {Filippini}\ and\ \citenamefont
  {Tasinato}(2019)}]{Filippini:2019cqk}%
  \BibitemOpen
  \bibfield  {author} {\bibinfo {author} {\bibfnamefont {F.}~\bibnamefont
  {Filippini}} and \bibinfo {author} {\bibfnamefont {G.}~\bibnamefont
  {Tasinato}},\ }\href {\doibase 10.1088/1361-6382/ab4371} {\bibfield
  {journal} {\bibinfo  {journal} {\emph {Class. Quant. Grav.}}\ }\textbf
  {\bibinfo {volume} {36}},\ \bibinfo {pages} {215015} (\bibinfo {year}
  {2019})},\ \Eprint {http://arxiv.org/abs/1903.02950} {arXiv:1903.02950
  [gr-qc]} \BibitemShut {NoStop}%
\bibitem [{\citenamefont {Fernandes}\ \emph {et~al.}(2019)\citenamefont
  {Fernandes}, \citenamefont {Herdeiro}, \citenamefont {Pombo}, \citenamefont
  {Radu},\ and\ \citenamefont {Sanchis-Gual}}]{Fernandes:2019kmh}%
  \BibitemOpen
  \bibfield  {author} {\bibinfo {author} {\bibfnamefont {P.~G.~S.}\
  \bibnamefont {Fernandes}}, \bibinfo {author} {\bibfnamefont {C.~A.~R.}\
  \bibnamefont {Herdeiro}}, \bibinfo {author} {\bibfnamefont {A.~M.}\
  \bibnamefont {Pombo}}, \bibinfo {author} {\bibfnamefont {E.}~\bibnamefont
  {Radu}},  and \bibinfo {author} {\bibfnamefont {N.}~\bibnamefont
  {Sanchis-Gual}},\ }\href {\doibase 10.1103/PhysRevD.100.084045} {\bibfield
  {journal} {\bibinfo  {journal} {\emph {Phys. Rev. D}}\ }\textbf {\bibinfo
  {volume} {100}},\ \bibinfo {pages} {084045} (\bibinfo {year} {2019})},\
  \Eprint {http://arxiv.org/abs/1908.00037} {arXiv:1908.00037 [gr-qc]}
  \BibitemShut {NoStop}%
\bibitem [{\citenamefont {Hawking}(1972)}]{Hawking:1972qk}%
  \BibitemOpen
  \bibfield  {author} {\bibinfo {author} {\bibfnamefont {S.~W.}\ \bibnamefont
  {Hawking}},\ }\href {\doibase 10.1007/BF01877518} {\bibfield  {journal}
  {\bibinfo  {journal} {\emph {Commun. Math. Phys.}}\ }\textbf {\bibinfo
  {volume} {25}},\ \bibinfo {pages} {167} (\bibinfo {year} {1972})}\BibitemShut
  {NoStop}%
\bibitem [{\citenamefont {Bekenstein}(1972)}]{Bekenstein:1972ny}%
  \BibitemOpen
  \bibfield  {author} {\bibinfo {author} {\bibfnamefont {J.~D.}\ \bibnamefont
  {Bekenstein}},\ }\href {\doibase 10.1103/PhysRevLett.28.452} {\bibfield
  {journal} {\bibinfo  {journal} {\emph {Phys. Rev. Lett.}}\ }\textbf {\bibinfo
  {volume} {28}},\ \bibinfo {pages} {452} (\bibinfo {year} {1972})}\BibitemShut
  {NoStop}%
\bibitem [{\citenamefont {Hu}\ \emph {et~al.}(2000)\citenamefont {Hu},
  \citenamefont {Barkana},\ and\ \citenamefont {Gruzinov}}]{Hu:2000ke}%
  \BibitemOpen
  \bibfield  {author} {\bibinfo {author} {\bibfnamefont {W.}~\bibnamefont
  {Hu}}, \bibinfo {author} {\bibfnamefont {R.}~\bibnamefont {Barkana}},  and
  \bibinfo {author} {\bibfnamefont {A.}~\bibnamefont {Gruzinov}},\ }\href
  {\doibase 10.1103/PhysRevLett.85.1158} {\bibfield  {journal} {\bibinfo
  {journal} {\emph {Phys. Rev. Lett.}}\ }\textbf {\bibinfo {volume} {85}},\
  \bibinfo {pages} {1158} (\bibinfo {year} {2000})},\ \Eprint
  {http://arxiv.org/abs/astro-ph/0003365} {arXiv:astro-ph/0003365} \BibitemShut
  {NoStop}%
\bibitem [{\citenamefont {O'Hare}(2020)}]{AxionLimits}%
  \BibitemOpen
  \bibfield  {author} {\bibinfo {author} {\bibfnamefont {C.}~\bibnamefont
  {O'Hare}},\ }\href {\doibase 10.5281/zenodo.3932430} {\enquote {\bibinfo
  {title} {cajohare/axionlimits: Axionlimits},}\ }\bibinfo {howpublished}
  {\url{https://cajohare.github.io/AxionLimits/}} (\bibinfo {year}
  {2020})\BibitemShut {NoStop}%
\bibitem [{\citenamefont {Regge}\ and\ \citenamefont
  {Wheeler}(1957)}]{Regge:1957td}%
  \BibitemOpen
  \bibfield  {author} {\bibinfo {author} {\bibfnamefont {T.}~\bibnamefont
  {Regge}} and \bibinfo {author} {\bibfnamefont {J.~A.}\ \bibnamefont
  {Wheeler}},\ }\href {\doibase 10.1103/PhysRev.108.1063} {\bibfield  {journal}
  {\bibinfo  {journal} {\emph {Phys. Rev.}}\ }\textbf {\bibinfo {volume}
  {108}},\ \bibinfo {pages} {1063} (\bibinfo {year} {1957})}\BibitemShut
  {NoStop}%
\bibitem [{\citenamefont {Zerilli}(1970{\natexlab{a}})}]{Zerilli:1970se}%
  \BibitemOpen
  \bibfield  {author} {\bibinfo {author} {\bibfnamefont {F.~J.}\ \bibnamefont
  {Zerilli}},\ }\href {\doibase 10.1103/PhysRevLett.24.737} {\bibfield
  {journal} {\bibinfo  {journal} {\emph {Phys. Rev. Lett.}}\ }\textbf {\bibinfo
  {volume} {24}},\ \bibinfo {pages} {737} (\bibinfo {year}
  {1970}{\natexlab{a}})}\BibitemShut {NoStop}%
\bibitem [{\citenamefont {Zerilli}(1970{\natexlab{b}})}]{Zerilli:1970wzz}%
  \BibitemOpen
  \bibfield  {author} {\bibinfo {author} {\bibfnamefont {F.~J.}\ \bibnamefont
  {Zerilli}},\ }\href {\doibase 10.1103/PhysRevD.2.2141} {\bibfield  {journal}
  {\bibinfo  {journal} {\emph {Phys. Rev. D}}\ }\textbf {\bibinfo {volume}
  {2}},\ \bibinfo {pages} {2141} (\bibinfo {year}
  {1970}{\natexlab{b}})}\BibitemShut {NoStop}%
\bibitem [{\citenamefont {Kobayashi}\ \emph {et~al.}(2012)\citenamefont
  {Kobayashi}, \citenamefont {Motohashi},\ and\ \citenamefont
  {Suyama}}]{Kobayashi:2012kh}%
  \BibitemOpen
  \bibfield  {author} {\bibinfo {author} {\bibfnamefont {T.}~\bibnamefont
  {Kobayashi}}, \bibinfo {author} {\bibfnamefont {H.}~\bibnamefont
  {Motohashi}},  and \bibinfo {author} {\bibfnamefont {T.}~\bibnamefont
  {Suyama}},\ }\href {\doibase 10.1103/PhysRevD.85.084025} {\bibfield
  {journal} {\bibinfo  {journal} {\emph {Phys. Rev. D}}\ }\textbf {\bibinfo
  {volume} {85}},\ \bibinfo {pages} {084025} (\bibinfo {year} {2012})},\
  \bibinfo {note} {[Erratum:
  \href{https://doi.org/10.1103/PhysRevD.96.109903}{{\it Phys. Rev. D} {\bf
  96}, 109903 (2017)}]},\ \Eprint {http://arxiv.org/abs/1202.4893}
  {arXiv:1202.4893 [gr-qc]} \BibitemShut {NoStop}%
%%CITATION = ARXIV:1202.4893;%%
\bibitem [{\citenamefont {Kobayashi}\ \emph {et~al.}(2014)\citenamefont
  {Kobayashi}, \citenamefont {Motohashi},\ and\ \citenamefont
  {Suyama}}]{Kobayashi:2014wsa}%
  \BibitemOpen
  \bibfield  {author} {\bibinfo {author} {\bibfnamefont {T.}~\bibnamefont
  {Kobayashi}}, \bibinfo {author} {\bibfnamefont {H.}~\bibnamefont
  {Motohashi}},  and \bibinfo {author} {\bibfnamefont {T.}~\bibnamefont
  {Suyama}},\ }\href {\doibase 10.1103/PhysRevD.89.084042} {\bibfield
  {journal} {\bibinfo  {journal} {\emph {Phys. Rev. D}}\ }\textbf {\bibinfo
  {volume} {89}},\ \bibinfo {pages} {084042} (\bibinfo {year} {2014})},\
  \Eprint {http://arxiv.org/abs/1402.6740} {arXiv:1402.6740 [gr-qc]}
  \BibitemShut {NoStop}%
%%CITATION = ARXIV:1402.6740;%%
\bibitem [{\citenamefont {Kase}\ and\ \citenamefont
  {Tsujikawa}(2022)}]{Kase:2021mix}%
  \BibitemOpen
  \bibfield  {author} {\bibinfo {author} {\bibfnamefont {R.}~\bibnamefont
  {Kase}} and \bibinfo {author} {\bibfnamefont {S.}~\bibnamefont {Tsujikawa}},\
  }\href {\doibase 10.1103/PhysRevD.105.024059} {\bibfield  {journal} {\bibinfo
   {journal} {\emph {Phys. Rev. D}}\ }\textbf {\bibinfo {volume} {105}},\
  \bibinfo {pages} {024059} (\bibinfo {year} {2022})},\ \Eprint
  {http://arxiv.org/abs/2110.12728} {arXiv:2110.12728 [gr-qc]} \BibitemShut
  {NoStop}%
\bibitem [{\citenamefont {Kase}\ and\ \citenamefont
  {Tsujikawa}(2023)}]{Kase:2023kvq}%
  \BibitemOpen
  \bibfield  {author} {\bibinfo {author} {\bibfnamefont {R.}~\bibnamefont
  {Kase}} and \bibinfo {author} {\bibfnamefont {S.}~\bibnamefont {Tsujikawa}},\
  }\href {\doibase 10.1103/PhysRevD.107.104045} {\bibfield  {journal} {\bibinfo
   {journal} {\emph {Phys. Rev. D}}\ }\textbf {\bibinfo {volume} {107}},\
  \bibinfo {pages} {104045} (\bibinfo {year} {2023})},\ \Eprint
  {http://arxiv.org/abs/2301.10362} {arXiv:2301.10362 [gr-qc]} \BibitemShut
  {NoStop}%
\bibitem [{\citenamefont {Burrage}\ \emph {et~al.}(2023)\citenamefont
  {Burrage}, \citenamefont {Fernandes}, \citenamefont {Brito},\ and\
  \citenamefont {Cardoso}}]{Burrage:2023zvk}%
  \BibitemOpen
  \bibfield  {author} {\bibinfo {author} {\bibfnamefont {C.}~\bibnamefont
  {Burrage}}, \bibinfo {author} {\bibfnamefont {P.~G.~S.}\ \bibnamefont
  {Fernandes}}, \bibinfo {author} {\bibfnamefont {R.}~\bibnamefont {Brito}},
  and \bibinfo {author} {\bibfnamefont {V.}~\bibnamefont {Cardoso}},\ }\href
  {\doibase 10.1088/1361-6382/acf9d6} {\bibfield  {journal} {\bibinfo
  {journal} {\emph {Class. Quant. Grav.}}\ }\textbf {\bibinfo {volume} {40}},\
  \bibinfo {pages} {205021} (\bibinfo {year} {2023})},\ \Eprint
  {http://arxiv.org/abs/2306.03662} {arXiv:2306.03662 [gr-qc]} \BibitemShut
  {NoStop}%
\bibitem [{\citenamefont {Kocherlakota}\ \emph {et~al.}(2021)\citenamefont
  {Kocherlakota} \emph {et~al.}}]{EventHorizonTelescope:2021dqv}%
  \BibitemOpen
  \bibfield  {author} {\bibinfo {author} {\bibfnamefont {P.}~\bibnamefont
  {Kocherlakota}} \emph {et~al.} (\bibinfo {collaboration} {Event Horizon
  Telescope}),\ }\href {\doibase 10.1103/PhysRevD.103.104047} {\bibfield
  {journal} {\bibinfo  {journal} {\emph {Phys. Rev. D}}\ }\textbf {\bibinfo
  {volume} {103}},\ \bibinfo {pages} {104047} (\bibinfo {year} {2021})},\
  \Eprint {http://arxiv.org/abs/2105.09343} {arXiv:2105.09343 [gr-qc]}
  \BibitemShut {NoStop}%
\end{thebibliography}%

\end{document}